# Hybrid Polymer-Garnet Materials for All-Solid-State Energy Storage Devices

Juan C. Verduzco, John N. Vergados, Alejandro Strachan, and Ernesto E. Marinero

School of Materials Engineering and Birck Nanotechnology Center
Purdue University, West Lafayette, Indiana, 47907 USA

**Abstract**

Hybrid electrolyte materials comprising polymer-ionic salt matrixes embedded with garnet particles constitute a promising class of materials for the realization of all-solid-state batteries. In addition to providing solutions to the safety issues inherent to current liquid electrolytes, hybrid polymer electrolytes offer advantages over other solid-state electrolytes. This is because their functional properties such as ionic conductivity, electrochemical stability, mechanical and thermal properties can be tailored to a particular application by independently optimizingthe properties of the constituent materials. Thereby, providing a rational approach to solving bottlenecks currently preventing solid-state electrolytes from practical implementation into battery devices. This review starts with a survey of solid-state electrolytes, focusing on their materials and ion transport limitations. Next, we summarize the current understanding oftransport mechanisms in composite polymer electrolytes (CPEs) with the purpose of identifyingmaterials solutions for further improving their properties. The overall goal of the review is to foster heightened research interest in these hybrid structures to rapidly advance developmentof future all-solid-state battery devices.

## 1  Introduction

Electric vehicles (EVs), internet-of-things (IoT) devices, wearable electronics and renewable energy generation are emerging technologies requiring significant improvements of current energy storage devices. These applications are evolving at an unprecedented rate and advanced materials are needed for the development of batteries affording greater capacity and energy density, higher performance, and fast charging capabilities. Presently, the most advanced rechargeable devices are Lithium-ion batteries (LIBs) and their global market is projected to reach 132 billion USD by 2025.LIBs provide the required ionic conductivity (IC) for commercial applications by using flammable liquid electrolytes. Thereby posing serious safety risks to both battery manufacturers and end-users including toxicity, flammability, and leakage. In addition, the liquid electrolyte's inherent chemical incompatibility with high energy anode materials, such as pure Lithium metal, limits their ultimate attainable energy density and capacity. Furthermore, rapid mass transport in these electrolytes generates deleterious concentration and polarization gradients hindering fast charging, a key requirement for the future of electrified transportation. The electrolyte flammability and the formation of lithium dendrites can result in thermal runaway events with catastrophic consequences. Dendrites form as ions are shuttled back and forth from the anode to the cathode during charge and discharge cycles. The Li-metal deposition is nucleated by topological variations in the anode and dendrites form in locations with lower impedance or higher localized electric fields.[1]

Thus, alternative battery materials are needed, and solid-state batteries (SSBs) offer a safe and promising solution to explore new electrochemistries for the realization of future advanced batteries. In SSBs, the liquid electrolytes are replaced by solid-phase materials. This readily alleviates safety concerns as solids present no leakage issues and the correct material choice can eliminate flammability and toxicity issues. Solid-state Electrolytes (SSE) exhibit excellent electrochemical stability, low reactivity with electrode materials such a Li-metal anodes and high voltage cathodes and exhibit desirable mechanical properties. However, significant issues that have prevented successful commercialization persist in SSBs, including the limited IC, the formation of dendritic structures, and the delamination at the anode and cathode electrode interfaces [2].

The implementation of Li-metal and multi-valence anode materials is considered strategic to increment the capacity and energy density of future rechargeable batteries. SSE materials span several material classes including oxides, sulfides, hydrides, halides, borates, phosphates, and polymers. These materials exhibit exceptional advantages against liquid electrolytes in certain attributes, but also significant shortcomings in other aspects that prevent their integration in battery devices. For example, chalcogenide glasses (sulfides) exhibit IC on par with those of liquid electrolytes; however, they are extremely hygroscopic and ionically conductive protective layers must be used in combination with them to protect them from ambient moisture [3].

Oxide ceramic garnets excel in chemical and electrochemical stability, mechanical strength, and electrochemical oxidation voltage stability. However, some of their notable disadvantages include brittleness, excessive fabrication costs, low IC ($10^{-5}$ - $10^{-3}$ S/cm),[3] and large Ohmic barriers at electrode interfaces. Additionally, despite their solid nature, lithium dendrites are known to nucleate at microstructural defects such as pores and grain boundaries.[4] Polymers, on the other hand, offer advantages such as stability against Li-metal, mechanical flexibility, low cost materials, and scalable manufacturing. However, their thermal stability, low oxidation voltages and inferior ionic conductivity ($10^{-7}$ - $10^{-5}$ S/cm)[3] are severe limitations for their implementation in battery devices. Polymer-based electrolytes are classified by their constituent materials into dry-solid polymer electrolytes (SPE), gel polymer electrolytes (GPE), and composite polymer electrolytes (CPE). SPE consist of a polymer host matrix with a complexed ionic salt dissociated into them. These systems as indicated, are limited by their inferior IC at room temperature. To compensate for this drawback, ionic liquids and plasticizers have been introduced to yield stable high ionic conductivity GPE. Unfortunately, the re-introduction of liquid components works to the detriment of mechanical properties, safety and reactivity to metal electrodes.[5]

CPE have gained significant interest as they provide the opportunity to combine the best attributes of polymer and oxide ceramic materials. The most promising CPE comprise oxide particles incorporated into a polymer-ionic salt matrix. The addition of filler particles into the polymer matrix results in significant improvements of its IC. Their development stems from the need to increase the dielectric constant of the host polymer to enhance the ionic salt dissociation.[5] Over the last few years our group has been developing thin film CPEs comprising PEO:LiTFSI matrixes embedded with aliovalently-substituted $Li_7La_3Zr_2O_{12}$ particles for their implementation in solid state batteries. We find that the Li-molar content of the garnet particle plays a critical role on the garnet particle % wt. load required to attain the highest IC in our CPEs. We ascribe this effect to polymer morphological changes induced by the particle physico-chemical properties resulting in formation of high conductivity channels in the polymer matrix [6]. However, the specific particle property that is modified by the Li-molar content and is responsible for the polymer morphology change remains to be identified.

A multiplicity of high dielectric constant filler particles has been added to various polymers with different degrees of success in improving the host polymer's IC. Of particular interest is the utilization of ionic-conducting ceramic particles to produce new hybrid materials with high dielectric constants and enhanced IC in comparison to the additions of non-ionically conducting filler particles.[7] Most promising are CPEs that employ low filler garnet particle loads,[6] thereby reducing the amount rare-earth elements and the materials cost of CPEs. The structure of this mini review is as follows: Section 2 summarizes salient studies on improving the ionic conductivity of CPEs. A brief description is also given of proposed transport mechanisms that are ascribed to increment IC in CPEs. In section 3, we discuss potential materials approaches to improve IC by morphological manipulation of the polymer matrix as well as optimization of the constituent materials. We conclude by providing our perspective on research areas that could significantly advance the realization of all-solid-state batteries based on CPEs.

## 2 Composite polymer electrolytes

Polyethylene oxide (PEO) is the most utilized ionically conducting polymer in CPEs and has gained technological relevance to energy storage applications. A comprehensive compilation of some of the current avenues of research regarding polymer electrolytes based in PEO can be found elsewhere [2]. Whereas the physical properties of PEO are not ideal for device applications, it has been widely employed as a model material to improve and understand ion transport in these hybrid materials. Some of PEO's inherent properties that are attractive for battery applications, include: mechanical flexibility, chemical and electrochemical stability against lithium metal anodes, low toxicity and its relatively high dielectric constant that enables remarkable solubility of lithium ionic salts in the polymer.[8] Important shortcomings of PEO for battery applications are: its low melting point ($\sim$ 60 °C), which falls within the expected operating temperature range of batteries, and its intrinsic IC, which is $\sim 10^4$ times lower at room temperature than that of liquid electrolytes.

Adding non-ionically conducting inorganic filler particles such as $Al_2O_3$, $MgO$, $TiO_2$ amongst others, as well as ionically conducting oxides such as the garnets $Li_7La_3Zr_2O_{12}$ (LLZO) to PEO:salt matrixes increments their IC substantially. Filler particle material properties, including particle size, surface chemistry and intrinsic IC have been shown to strongly influence ionic transport in CPEs. The amount of filler particles necessary to increase the IC of PEO-based CPEs has been found to depend, in addition to the nature of the filler material, on the EO:Li ratio (ratio of ether oxygen groups in the polymer chain to the amount of Li ions provided by the solvated Li-ionic salts), and on the polymer molecular weight (MW). The filler amounts needed to optimize IC have been reported for different filler materials to range from 5% to 52% weight load.[6] Nevertheless, to-date the highest room temperature IC values reported in CPEs are around $10^{-4}$ - $10^{-3}$ S/cm[7] and remain below that of liquid electrolytes ($10^{-2}$ S/cm). Transference numbers, the ratio of the electric current derived from the cation transport to the total current, is another property of SSBs that can be improved by the implementation on CPEs. Transference numbers close to unity, which are highly desirable for transport processes, are possible in CPEs; however, competing diffusion processes between the larger anion salts and the small Li-cations result in substantially lower transference numbers. In addition, the limited solvation of ionic salts in CPEs reduces the concentration of mobile ions available for transport. Inadequate interfacial contact between CPE and the electrode materials, caused by poor wettability between the solid materials, also constitutes a significant challenge.[9] Interfacial contact is affected by the roughness and topology of the surfaces. Poor interfacial contact inhibits ionic transport, hinders adhesion, and compromises mechanical properties. Finally, the formation of new phases at interfaces introduces charge transfer resistance, thereby reducing charge/discharge efficiency and cyclability.[4] The electrode expansion and shrinkage during each cycle serves to further compound the loss of contact and can ultimately result in delamination and materials failure.[10]

## 2.1 Ion transport in composite polymer electrolytes

The main mechanisms proposed to explain ionic transport in CPEs are: (i) filler particles decrease the polymer crystallinity, increment the amorphous fraction, and improve the Li$^+$ mobility due to higher amorphous phase segmental motions; (ii) filler particles through Lewis acid-base interactions adsorb anions, thereby breaking up the ion pairs, leading to increased dissociation of lithium salts, and thus increasing the Li$^+$ ion concentration. Inorganic compounds that are acidic or neutral are more likely to form hydrogen bonding with the salt anions and with the oxygen in PEO, potentially promoting efficient salt dissociation and faster Li-ion transport;[11] (iii) the interaction between the filler particles and the surrounding polymer structure creates microstructural "highways" for efficient lithium ion transport. These are particularly effective when said highways become interconnected across the bulk polymer matrix. Noteworthy in this respect is the particle size and its aspect ratio; (iv) in CPEs with large particle weight loads, particle percolation provides additional channels for transport; (v) interface contributions: formation of space-charge layers facilitates ion transport at the filler particle-polymer boundaries; (vi) the ion mobility and ion transference number are intrinsically determined by the lithium salt and the complexation between the salt and the polymer matrix. Additions of passive fillers ($Al_2O_3$, $SiO_2$, $TiO_2$) to PEO:$LiClO_4$ improves ionic conductivity through weakening the polyether oxygen-Li interactions [12].

As mentioned, ionic transport within the polymer matrix has been linked to the segmental motion of polymer chains. Polymer matrixes complexed with dissociated salts are usually described by the Vogel–Tamman–Fulcher (VTF) equation with parameters obtained from free volume and configurational entropy [2]. Ether-oxygen units disassociate ionic lithium salts and coordinate with the resulting free lithium ions. Ionic transport results from a combination of intrachain and interchain ion hopping as a sequential process of creating and breaking lithium-oxygen bonds as illustrated in Fig. 1.[8]

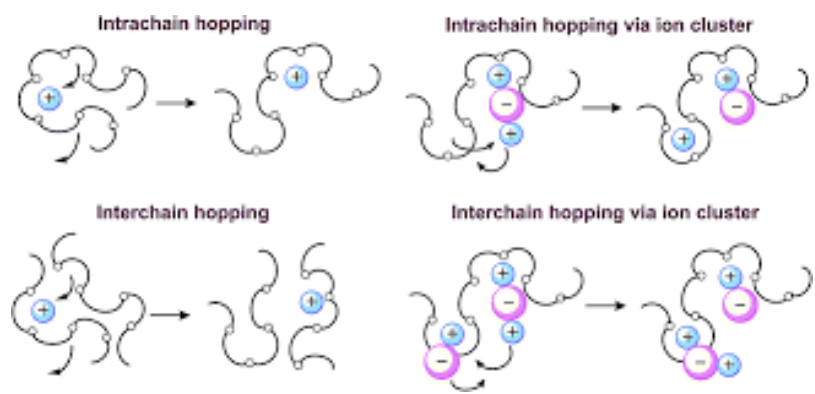

Figure 1: Proposed hopping mechanism for ion transport in PEO. Reprinted with permission from Ref.[8] Copyright 2015. Journal of Materials Chemistry A.

Ionic conductivity has been suggested to occur predominantly in PEO-based CPEs through the polymer amorphous regions, as the crystalline domains restrict the motion of the polymer chains, thereby inhibiting ionic conductivity. Molecular dynamics simulations with external applied fields [13] provide a picture of transport with unparalleled resolution and are providing insight into the molecular level processes involved [14]. Atomic level studies of the link between correlated segmental relaxation in the polymer[15] with ionic hopping could provide definite mechanistic answers.

Ionic conductivity can also be optimized in semi-crystalline polymers by altering the amount of lithium salts dissociated in the polymer [16], introduction of plasticizers [17], and through mechanical strain-deformation of the polymer [18]. Fullerton and Maranas[16] attributed improvements of IC to changes within localized regions of amorphous material that form between crystalline regions and lamellae. Chen et al.[17] reported that the intensities of the XRD peaks related to crystalline PEO were significantly reduced with increasing addition of Succinonitrile plasticizer. Kelly et al.[18] used Polarization Light Microscopy (PLM) to characterize films before and after a tensile strain deformation through stretching. They observed the growth of amorphous regions, to which they attributed the IC improvements. The role that the microstructure plays on IC has led to promising approaches for the design and synthesis of microstructures conducive to high ionic transport. This includes the utilization of block copolymers to achieve directional ion transport and reduce the amount of scattering through lesser conductive regions.[19]

## 2.2 Ion mobility in CPEs - Lewis Acidity

Polymers for SSEs are chosen for their high ionic salt solubility and lower anion diffusivity relative to the lithium ions. Studies and simulations have shown that increasing the Lewis acidity of polymers can increase IC. Fig. 2 from the work of Savoie et al.[20] presents the molecular basis for preferential anion diffusion in PEO when compared to a series of polyborane polymers. PEO is more acidic than the other polymers and strong coordination of $Li^+$ in PEO is reflected in the helical distortion of the polymer structure about the ion. Ion transport in the polymer is controlled by the magnitude of the activation energy that binds the Li ions to the reactive sites in the polymer chain, such as ether-oxygen sites in PEO. This binding energy is influenced by the presence of acid or basic Lewis centers at the filler particle surface. The nature and strength of these interactions and their mechanistic effect on the transport mechanism is still not fully understood. Dissanayake et al.[21] suggest that ionic conductivity within the filler particles is not responsible for transport enhancements in CPEs, but rather, that it is the interactions between the particle surface groups and the Li ions that drive the enhancement. At low filler contents, particles give rise to favorable conducting paths in the vicinity of their surface; in contrast, at high filler contents, segmental relaxation of the polymer chains is immobilized by the increased number of particles. In their studies of the $(PEO)_9$ - LiTFSI - $Al_2O_3$ system, with the filler particles functionalized to exhibit either acid, neutral or basic characteristics, it is the CPE embedded with particles with acidic nature that display the highest ionic conductivity. Lewis basic polymers, like PEO, tend to form strong coordination bonds with cations and restrict their mobility. Increasing the acidity of a polymer, allows cations to move more freely in the matrix while anion mobility is restricted. Several studies have been conducted on the effects of increasing the Lewis acidity of ionically conductive polymers. For example, Matsumi et al. obtained ionic conductivity in organoboron polymer electrolytes of $3.05 \times 10^{-5}$ S/cm. Research in this area could lead to significant improvements in ionic conductivity in this material class.[22]

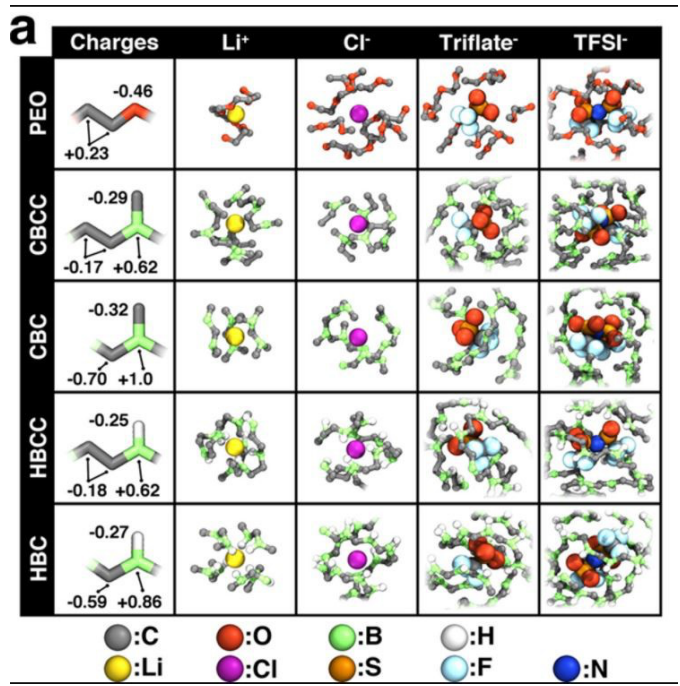

Figure 2: Ion coordination behavior in Lewis-basic and Lewis-acidic polymers. Reprinted with permission from Ref.[20] Copyright 2017. Journal of Physical Chemistry Letters.

## 2.3 Fillers and plasticizers

Incorporation of diverse high ionically conducting ceramic fillers in CPEs has been successful in improving their room temperature IC. The specific properties of these fillers that drive the IC enhancement are actively under investigation and the outcomes will be extremely beneficial for the design of novel CPEs. The influence of filler particle size was studied by Capiglia et al. through the addition of $BaTiO_3$ with particle sizes ranging from 0.5 μm to 60 nm.[23] Their results showed that the smaller filler particles influenced the polymer morphology and that the larger ones primarily improved the solvation of the lithium salt. The role of surface charge in passive fillers was studied by Dissanayake et al. to test the Lewis-acid/base conductivity hypothesis. They concluded that the active H/OH sites on the surface of the fillers drive the ionic conductivity improvement.[21]

Croce et al.[24] proposed a model that attributes increase in IC of CPEs to two factors: a) the effect of the fillers on PEO structural modifications that result in the formation of $Li^+$ conducting pathways at the filler particle surface; b) the interaction of filler particles with ionic salts promoting their dissociation. This study attributed transport enhancements in PEO-based composites to interactions between particle surface groups and both, the PEO segments and the electrolyte ionic species. The locally induced modifications result in an increase of the fraction of free $Li^+$ ions which can move fast throughout the conducting pathways at the ceramic extended surface. This was further studied by Tambelli et al. in their experiments using $α$-$Al_2O_3$ and $γ$-$Al_2O_3$ filler particles. They suggested the existence of a space-charge region around the filler particle that is affected by the surface charge of the particle, thereby influencing the IC of the composites.[25] LATP($Li_{1.3}Al_{0.3}Ti_{1.7}(PO_4)_3$) as a filler in PEO-$LiClO_4$ membranes has been investigated by Ban et al. [26]. They report high IC at T > 50°C in composites loaded with 50% wt. LATP nanoparticles. The membranes exhibited attractive mechanical and electrochemical properties.

The introduction of solid plasticizers has also been studied with moderate success in increasing room temperature IC. Different plasticizers were used in the studies of the PEO-LiClO$_4$ system and improvements in IC were observed to depend on the value of the dielectric constant of the plasticizer material.[27] A combination of both ceramic fillers and plasticizer was studied by Chen et al.[17] They found that the addition of solid plasticizer Succinonitrile (SN) resulted in one order of magnitude improvement in IC of the PEO-LiTFSI-LLZO system. The highest reported IC value in their work reached 1.9 x10$^{-4}$ S/cm. Their results are shown in Fig. 3 in which the ionic conductivity temperature dependence as a function of added plasticizer is plotted. The insert in Fig. 3 displays the conductivity dependence on LLZO % wt at 25°C. The increase in IC with LLZO additions is ascribed by the authors to the particles limiting the movement of anions. A battery arrangement utilizing the electrolyte PEO-LiTFSI-LLZO + 10% SN in between a LiFePO$_4$ (LFP) cathode material and lithium metal anode was tested. Cyclability studies at 1 C rate showed a specific capacity of 108.8 mAhg$^{-1}$ and 80.0% retention rate after 500 cycles with negligible change in interfacial resistance. This highlights the importance of a stable electrolyte-electrode interface for extended battery cycling.

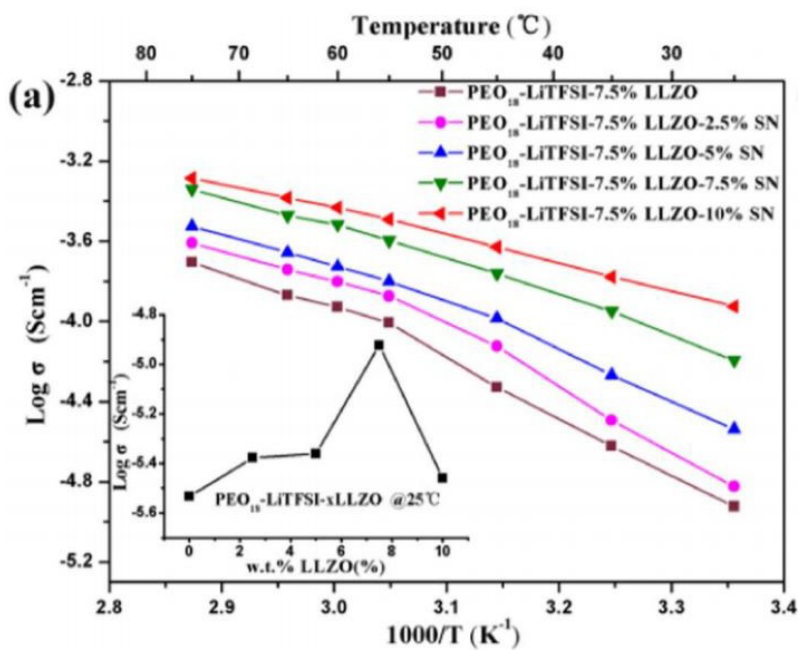

Figure 3: Lithium ion conductivities of CPE films with increasing content of SN plasticizer. Reprinted with permission from Ref.[17] Copyright 2018. Journal of the Electrochemical Society.

# 3 Structurally organized materials for CPEs

## 3.1 Self-organized polymer structures

Novel polymer structures based on self-assembled block copolymer electrolytes (BCEs) have been investigated with promising results. Block copolymers comprise chemically dissimilar polymer segments and offer the possibility of creating high ionic conductivity microdomains separated from the main mechanical scaffold created by the second component. Moreover, these channels can be oriented in preferential directions to facilitate transport. For example, Majewski et al. showed a 10-fold increase in IC of magnetically aligned BCEs based on PEO with channels that were oriented perpendicular to the surface of the electrodes. Their results are presented in Fig. 4 [28].

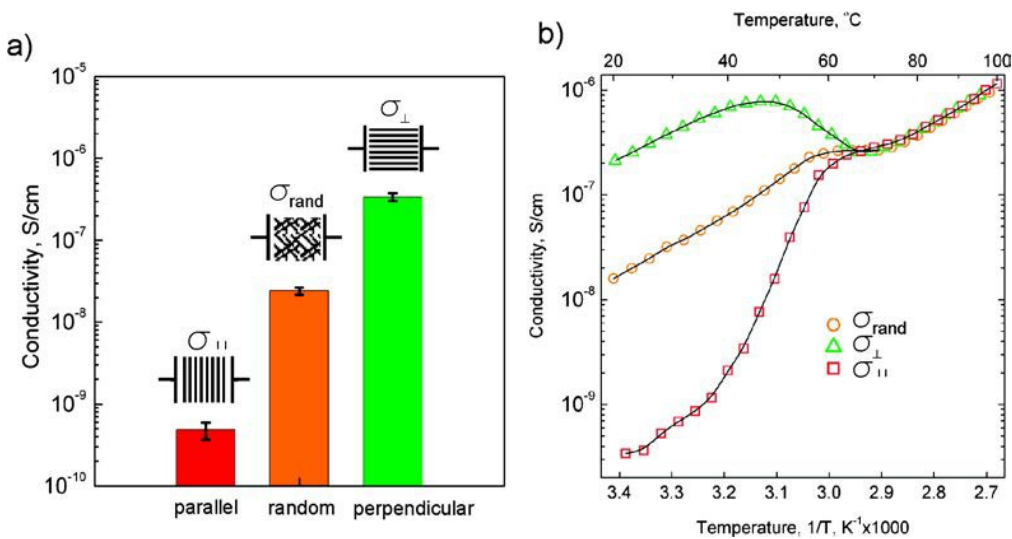

Figure 4: (a) Directional room temperature conductivies for the PEO-b-PMA/CB system and (b) Temperature dependence of the ionic conductivity. Reprinted with permission from Ref.[28] Copyright 2010. Journal of the American Chemical Society.

Gomez et al. studied a system based on PS-PEO block copolymers and the ionic salt LiTFSI. They reported an enhancement in ionic conductivity with increasing molecular weight of the copolymers, but also that the distribution of ions, shown in Fig. 5, from the ionic salt dissociation occurs preferentially within the PEO microdomain lamellar segment of the BCE. The selective salt dissociation is attributed to ion coordination with the PEO ether oxygen groups and to the existence of non-uniform stress fields.[19]

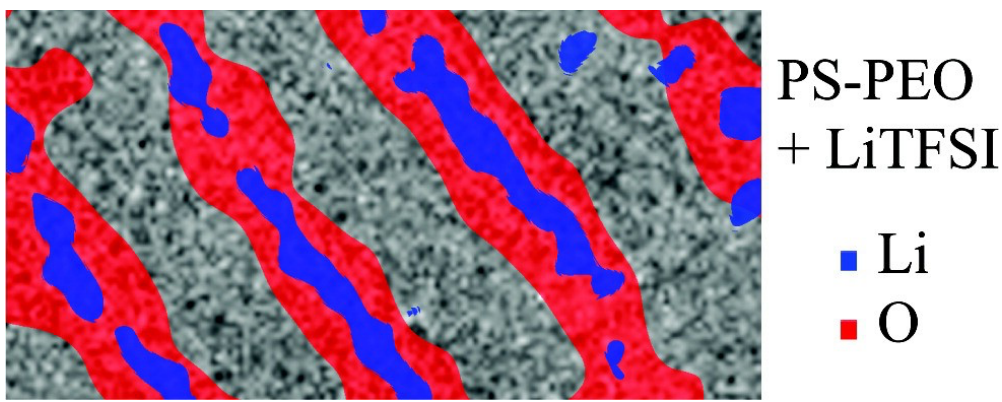

Figure 5: Cation distribution in a PS-PEO block copolymer. Reprinted with permission from Ref.[19] Copyright 2009. Nano Letters.

## 3.2 Chemical manipulation of the polymer morphology

Whereas significant efforts have been devoted to the improvement of the intrinsic IC of LLZO-type garnet materials, virtually no reports exist on how the garnet particle physico-chemical properties influence the IC of CPEs. For example, aliovalent substitution is extensively used to stabilize the high IC cubic LLZO garnet polymorph and different dopants have been employed to increment IC. The effect of dopants on the structural and transport properties of these garnet materials have been characterized with a multiplicity of techniques including SEM, TEM, XRD, NMR, NPD (Neutron Powder Diffraction), EIS as well as in-situ operando XRD techniques. In addition, the work of Lu et al. [29] indicates that using Fluorine as a dopant improves the IC of the garnet-type $Li_{6.25}Ga_{0.25}La_3Zr_2O_{12}$ (LGLZO).

Nonetheless, optimization of the IC of the ceramic filler itself doesn't directly correlate with an improvement of the IC of the CPE material hosting it, as peaks for CPEs IC are often found at low particle %wt loads wherein ion transport through the garnet particles does not play a significant role on the IC of the CPE. Thus, other properties of the particles must be responsible for the IC enhancement. We have recently investigated the dependence of the IC of PEO:LiTSFI matrixes embedded with Bi-doped garnet particles as a function %wt. load and garnet Li-molar content. It is noted that Bi-substitutions into the Zr-site modifies the Li-molar content to maintain charge neutrality.[6] The sol-gel Pechini method was employed to synthesize garnet particles with nominal compositions: Bi-LLZO ($Li_6La_3Zr_1Bi_1O_{12}$), 0.75Bi-LLZO ($Li_{6.25}La_3Zr_{1.25}Bi_{0.75}O_{12}$) and 0.75BiNd-LLZO ($Li_{6.25}La_{2.8}Nd_{0.2}Zr_{1.25}Bi_{0.75}O_{12}$).

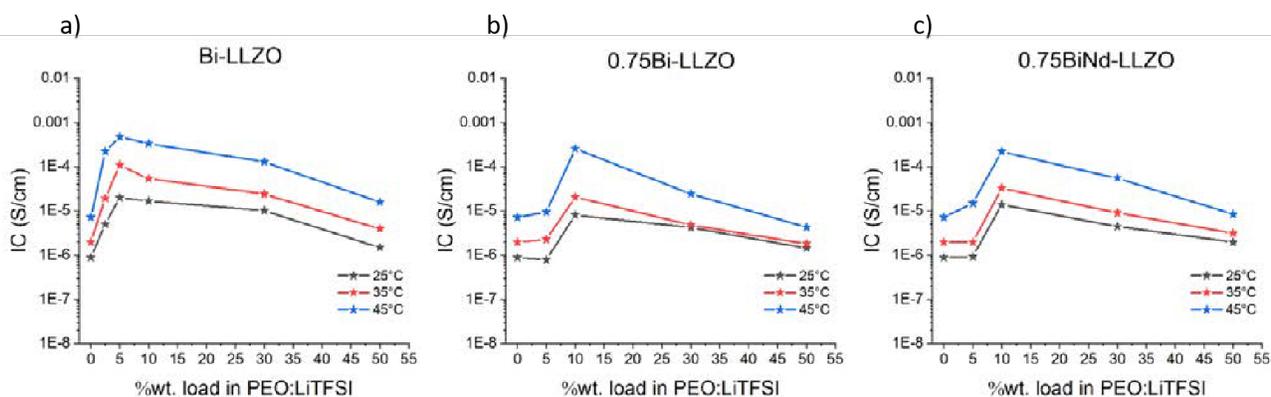

Figure 6: Ionic conductivity of PEO:LiFTSI matrixes vs. garnet particle %wt. load and temperature for (a) Bi-LLZO ($Li_6La_3Zr_1Bi_1O_{12}$), (b) 0.75Bi-LLZO ($Li_{6.25}La_3Zr_{1.25}Bi_{0.75}O_{12}$) and (c) 0.75BiNd-LLZO ($Li_{6.25}La_{2.8}Nd_{0.2}Zr_{1.25}Bi_{0.75}O_{12}$). Note the Li-molar dependence on particle %wt. load required for both deriving the optimum IC and its magnitude[6]

Measurements are presented in Fig. 6 and they indicate that: a) the IC maximizes for low %wt. loads (5-10%); b) the %wt. load required for highest IC and its magnitude depend on the Li-molar content; c) the Li-molar content dominates over structural or electronic modifications induced by incorporating a second dopant at the La-site.

We ascribe the enhancement of the IC in this low particle %wt. loaded CPEs and the dependence on the garnet Li-molar content to the formation of a polymer morphology comprising a network of interconnected amorphous regions between neighboring spherulites that are heterogeneously nucleated by the garnet particles. The garnet Li-molar content is suggested to control the particle surface properties and thereby, the nucleation and growth of spherulites. Altering the garnet par- ticle Li molar content, changes the spherulite nucleation and growth and in turn, the %wt. load needed to form the optimum polymer morphology to facilitate macroscopic ion transport. This *chemical manipulation* of polymer morphology in hybrid composite polymer electrolytes presents an attractive approach to enhance IC in CPEs in conjunction with improvements in constituent material properties

# 4 Interfacial resistance and adhesion

The properties of the SSE and the battery electrode interfaces determine its ionic resistance and ion permeability as well as its cyclability, robustness, and performance. The formation of high ionic resistance electrode interfaces in SSBs constitutes a major bottleneck for their commercial implementation. Upon cycling, voids form at the anode/SSE interface due to volume expansion and pulverization of electrode particles.[10] This dramatically reduces battery life, power, efficiency, charging ability and cycle life, in particular at low temperatures. Several strategies have been employed to mitigate the electrode-electrolyte contact issues due to interphase formation. They include cathode- coatings, novel anode materials, and buffer layers.

Chi et al.[30] improved the interfacial contact in a PEO-garnet-LLZO CPE by intercalating an 8 $\mu$m thick soft PEO layers at the anode and cathode interfaces. The interlayer was a mixture of PEO (MW 100000) and the ionic salt LiTFSI was solvated with Acetonitrile. This improved the contact between the voids in the garnet and the electrode interfaces as illustrated in Fig. 7.

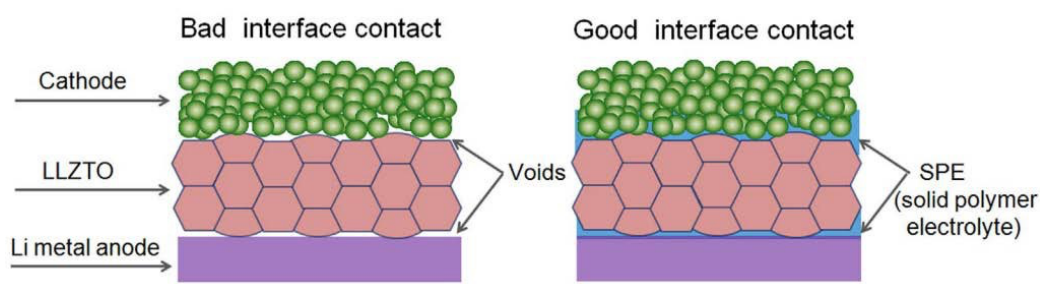

Figure 7: Schematics depicting poor interfacial contact due to garnet voids which is improved by the intercalation of SPE buffer layers between the SSE and the electrode materials. Reprinted with permission from Ref.[30] Copyright 2019. Energy Storage Materials.

Similarly, Yang et al.[31] introduced Polypropylene Carbonate-Based buffer layers to stabilize the interface of LIBs against a PEO solid polymer electrolyte. Their work showed that the introduction of this layer stabilized the battery Coulombic efficiency and maintained good interfacial contact during battery cycling. At operating temperatures of 50 °C, the buffer layer exhibited liquid-like properties that allowed for better contact with the anode surface after cycling. These methods to improve contact between solid electrode components are promising; however, the low sheer moduli of the interlayers enable dendrite formation and their low ionic conductivity negatively impacts transport. A potential solution is to utilize functionally engineered CPEs in multilayered and/or surface-modified structures that provide high bulk IC together with interfaces that provide the necessary adhesion, mechanical and transport requirements. Hybrid polymer- ceramic materials offer superior mechanical strength and can endure volume and stress changes upon cycling.[32]

# 5 Conclusions

Composite Polymer Electrolytes are one of the most promising solid electrolytes for the realization of commercial solid-state batteries. This is on account of their mechanical properties, electrochemical stability against electrode materials and their inherent safety attributes. Furthermore, the constituent materials and their fabrication are inherently low cost and are readily scalable. They can be expected to help reduce the cost of transport electrification and the expansion of renewable energy sources. Nevertheless, significant research and development efforts and advanced materi- als development are needed to circumvent critical bottlenecks facing CPEs, namely: 1) the ionic conductivity needs to be significantly improved, in particular at or below room temperature; 2) new materials and ion transport mechanisms need to be identified to satisfy the wide temperature operating range of battery devices; 3) materials solutions are needed to yield mechanically robust, low resistance CPE-electrode interfaces; 4) hybrid materials need to be identified whose properties meet the requirements for fast charging operation for EV applications.

Examination of the factors that influence ion transport in CPEs leads us to suggest focus areas of research and materials engineering development critical for their implementation into SSBs, namely: a) identify polymers, filler particles and anions salt materials whose physical and chemical properties synergistically interact to enhance ion transport; b) investigate hybrid polymers exhibiting self-assembly such as BCP as well as polymer blends, selected based on their individual attributes (mechanical strength, high IC, thermal properties) to provide the required CPE functionality; c) identify layered hybrid composites to provide low resistance, stable electrode interfaces, whilst the bulk of the solid electrolyte (employing different material architectures) provides the fast ionic conductivity; d) design and synthesis of new hybrid composite materials for fast charging applications. Furthermore, the development of high transference number CPEs as well as engineering solutions for thermal dissipation will be critical.

Future research that involves the creation of systems with novel filler materials and plasticizers can help alleviate the shortcomings currently experienced by CPEs. The creation of directional transport avenues through modifications of the polymer morphology and the tuning of ceramic filler properties to test their effect in the polymer host are both exciting avenues for research. Moreover, there is still a need for studies that can identify a polymer matrix that can perform at the operating temperatures of battery systems without compromising its mechanical properties and without allowing for the creation of dendritic structures. In summary, hybrid materials for CPEs provide a platform for the rational design of future all-solid-state batteries that can potentially solve current issues with solid electrolytes and pave the way for their integration into all-solid-state batteries comprising advanced anode and cathode materials and to exploit new battery electrochemistries.

# 6 Acknowledgements

J. C. V. thanks the Science and Technology Council of México (Consejo Nacional de Ciencia y Tecnología, CONACYT) for partial financial support of this research. We acknowledge the following undergraduate students at Purdue University for meaningful conversations regarding this research: David Starr and Julia K. Peck. Fruitful discussions with Professor Jeffrey P. Youngblood at Purdue University, and Dr. Andres Villa, are also acknowledged.

# 7 Author Information

**Corresponding Author**
**Ernesto E. Marinero** - School of Materials Engineering and Birck Nanotechnology Center, Purdue University, West Lafayette, Indiana 47907; https://orcid.org/0000-0001-5316-9342; Email: eemarinero@purdue.edu

**For Table of Contents Only - Graphic**

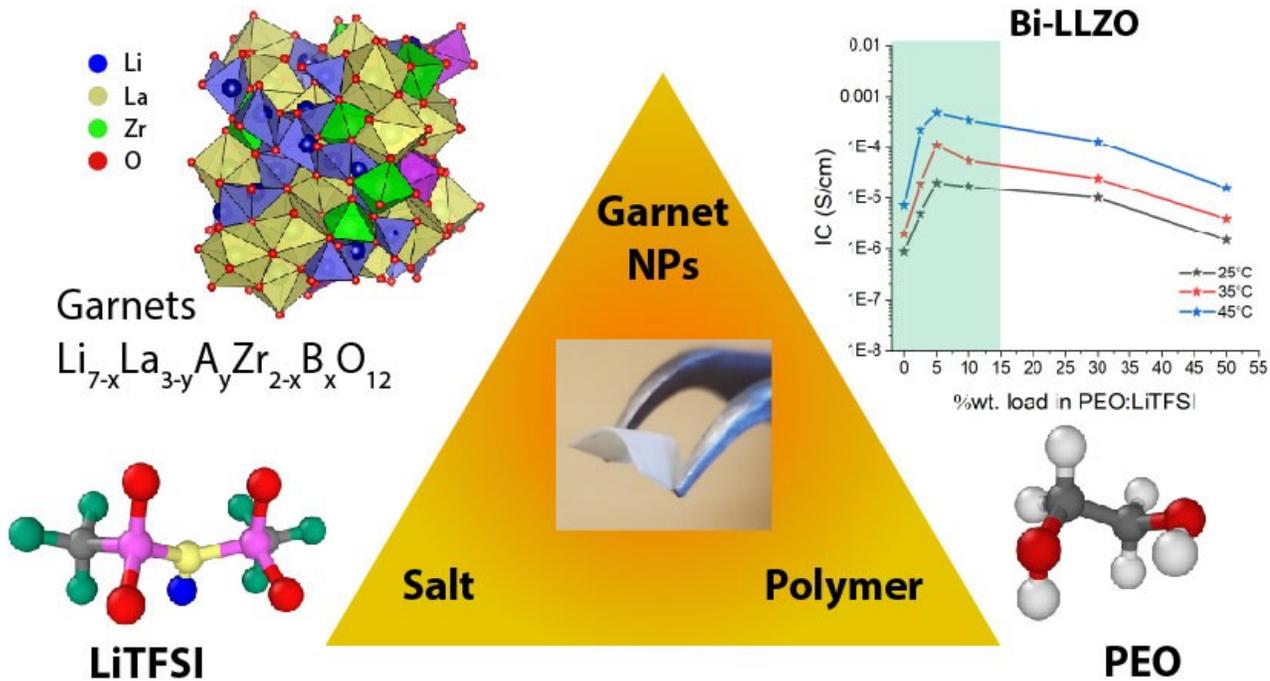